\documentclass[aps,twocolumn,superscriptaddress]{revtex4}
\usepackage{amsmath,amssymb,mathrsfs}
\usepackage{graphicx}
\usepackage{hyperref}
\usepackage{paralist}

\usepackage{subfigure}
\usepackage{dcolumn}
\usepackage{color}
\usepackage{bbold}

\def\be{\begin{equation}}
\def\ee{\end{equation}}
\def\bea{\begin{eqnarray}}
\def\eea{\end{eqnarray}}
\def\nn{\nonumber}

\def\tbf{\textbf}
\def\up{\uparrow} 
\def\down{\downarrow}

\def\npsection#1{{\par{\vskip 5pt}\noindent\bf #1}}

\begin{document}

\def\papertitle{Chiral spin superfluidity and spontaneous spin Hall effect of interacting bosons}
\title{\papertitle}
\author{Xiaopeng Li}
\affiliation{Condensed Matter Theory Center and Joint Quantum Institute, Department of Physics, 
University of Maryland, College Park, MD 20742-4111, USA}

\author{Stefan S. Natu} 
\affiliation{Condensed Matter Theory Center and Joint Quantum Institute, Department of Physics, 
University of Maryland, College Park, MD 20742-4111, USA}

\author{Arun Paramekanti} 
\affiliation{Department of Physics, University of Toronto, Toronto, Ontario M5S 1A7, Canada }
\affiliation{Canadian Institute for Advanced Research, Toronto, Ontario M5G 1Z8, Canada}

\author{S. Das Sarma}
\affiliation{Condensed Matter Theory Center and Joint Quantum Institute, Department of Physics, 
University of Maryland, College Park, MD 20742-4111, USA}

\begin{abstract}
Recent experiments on ultracold atoms in optical lattices have synthesized a variety of tunable bands with degenerate double-well
structures in momentum space. Such degeneracies in the single particle spectrum strongly enhance quantum
fluctuations, and may lead to exotic many-body ground states. Here we consider weakly interacting spinor Bose gases in 
such bands, and discover a universal quantum ``order by disorder''  phenomenon which selects a novel chiral spin superfluid with 
remarkable properties such as spontaneous anomalous spin Hall effect and momentum space 
antiferromagnetism. For bosons in the excited Dirac band of a hexagonal lattice, such a state supports 
staggered spin loop currents in real space. We show that Bloch oscillations provide a powerful dynamical 
route to quantum state preparation of such a chiral spin superfluid. Our predictions can be  readily tested in spin resolved time-of-flight experiments.
\end{abstract}

\maketitle 
The ability to optically address and manipulate the 
spin and momentum of electrons in a solid forms the basis for the fertile fields of spintronics and 
valleytronics~\cite{2004_DasSarma_RMP,2007_Xiao_PRL}.
Recent experimental progress in the field of ultracold atomic gases has led to the creation of 
optical lattices supporting bandstructures with multiple
minima (valleys)~\cite{2002_StamperKurn_PRL,
2009_Lin_Nat,2011_Lin_Spielman_Nature,2011_Hemmerich_NatPhys,2012_Chen_Pan_PRL,2012_Zhang_PRL,2012_Zwierlein_PRL,
2012_Sengstock_NatPhys,2012_Esslinger_Nature,2013_Chin_NatPhys,2013_Sengstock_NatPhys,
2013_Hemmerich_NJP,2013_Bloch_PRL,2013_Wolfgang_PRL,2014_Bloch_arXiv}, 
and setups which allow for a study of low-dimensional transport 
phenomena~\cite{2007_Esslinger_PRL,2012_Esslinger_Nature,2013_Chin_arXiv,2013_Spielman_Nature,2014_Esslinger_arXiv}. 
This valley degeneracy is achieved in experiments by considering 
atoms with Raman induced synthetic spin-orbit coupling~\cite{2013_Galitski_Nat,
2002_StamperKurn_PRL,2009_Lin_Nat,2011_Lin_Spielman_Nature,2012_Chen_Pan_PRL,2012_Zhang_PRL,2012_Zwierlein_PRL}, atoms in
shaken optical lattices~\cite{2013_Chin_NatPhys,2013_Sengstock_NatPhys}, and atoms loaded
into excited optical lattice bands~\cite{2011_Hemmerich_NatPhys,2012_Sengstock_NatPhys,
2013_Hemmerich_NJP,2012_Esslinger_Nature,2011_Liu_NatPhys} or engineered  $\pi$-flux 
lattices~\cite{2013_Bloch_PRL,2013_Wolfgang_PRL,2014_Bloch_arXiv}. 
These landmark developments herald the emergence of valleytronics (or atomtronics) for cold atoms, and
set the stage for the discovery of novel phases of atomic matter. 

The presence of multiple valley and spin degrees of freedom often leads to a large degeneracy of single-particle
ground states. When such extensive degeneracies persist at mean field level, many-body fluctuations
play a crucial role in selecting the eventual ground state. Indeed, this is the 
basis for fascinating phases such as fractional quantum 
Hall liquids in degenerate Landau levels~\cite{1982_Tsui_PRL}, 
unexpected magnetic orders in quasi-one dimensional bands~\cite{2013_Chen_PRL,2013_Wu_arXiv}, 
and highly entangled quantum spin liquids in frustrated magnets~\cite{2010_Balents_Nature}. 
In certain systems, fluctuations can select unusual long-range ordered
many-body states which have the maximal entropy out of the set of energetically degenerate states, 
a phenomenon dubbed `thermal order by disorder'~\cite{1980_Villain_JPF,1989_Henley_PRL}.
On the other hand, at low temperatures, the selection may favor ordered states with lower quantum zero point fluctuation on top of
the mean field energy, leading to `quantum order by disorder'~\cite{1989_Henley_PRL}. 
A direct identification of this phenomenon in solid state systems is, however, often
complicated by the presence of ordinarily negligible and material-specific terms in the Hamiltonian which can
overwhelm the order-by-disorder physics. 
Ultracold atoms, with clean and well-characterized tunable Hamiltonians, 
provide a particularly attractive platform to expose this remarkable phenomenon.

Single species of repulsive bosons loaded into a multivalley dispersion will typically condense
at a single minimum, due to mean-field interactions. This spontaneously broken valley symmetry 
concurrently leads to a broken inversion and time-reversal symmetry (TRS). Such a condensate in a $\pi$-flux triangular
lattice yields staggered charge loop current order~\cite{2012_Paramekanti_PRA,
2012_You_PRL,2013_Paramekanti_PRB,2014_Zaletel_PRB,2014_Bloch_arXiv} on triangular 
plaquettes. For weak interactions, the physics of this state is well captured by 
Gross-Pitaevskii theory~\cite{2008_Pethick_Smith}. By contrast, as we show here, the physics of multi-component 
bosons loaded into such bands is far richer as the spin and valley degeneracies 
persist even at the classical interacting level, and quantum fluctuations play a crucial role 
in selecting an exotic ground state.

Here we study two component or equivalently pseudo-spin-$1/2$ bosons 
loaded into such a multivalley band. 
This leads to an extra spin-valley degeneracy since
each spin state can be localized in one of two valleys. We show that quantum fluctuations lead to a `quantum order by
disorder' effect in such a system, where opposite spins condense at the two minima, giving rise to 
chiral spin order in the system. Remarkably, this selection is ``{\it universal}" in that it 
is independent of the microscopic details such as the lattice geometry or the precise dispersion, 
and is guaranteed by the symmetry which protects the valley degeneracy. 
The most direct experimental consequence of this chiral spin order is 
$\int d^d \tbf{k} \, \tbf{k} \left[ n_\up( \tbf{k}) - n_\down(\tbf{k})\right] \neq 0 $ 
and 
$\int d^d \tbf{k}  \left[ n_\up( \tbf{k}) - n_\down(\tbf{k})\right] = 0 $,  
 with $n_{\up/\down} (\tbf{k})$  the spin resolved momentum distribution. 
The emergent  coupling between spin and orbital motions leads to interaction-induced spontaneous spin Hall effect of bosons 
in optical lattices lacking inversion symmetry.

Taking a concrete example of spinor bosons loaded at 
massive Dirac points of a graphene-like lattice, such as that recently realized experimentally~\cite{2012_Sengstock_NatPhys}, 
we predict that chiral spin order implies spin loop currents in real space. 
With increasing interaction strength, we find a rich phase diagram with phase transitions into 
partially or fully spin polarized superfluid states and Mott states together with an emergent 
quantum tricritical point. We show that Bloch oscillation techniques provide a high fidelity 
route to preparing the chiral spin superfluid and studying the concomitant bosonic spin Hall phenomena.

\begin{figure}[htp]
\includegraphics[angle=0,width=\linewidth]{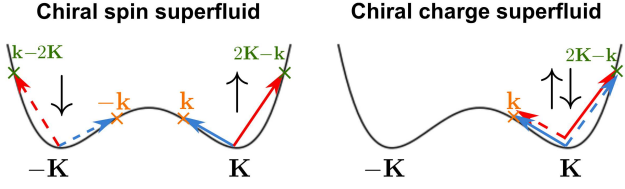}
\caption{Lowest order scattering processes in chiral spin and charge superfluid states in the 
``double-valley'' band. 
In the chiral spin state $(+-)$, 
two condensate atoms of opposite spin at ${\bf K}$ and $-{\bf K}$ are 
scattered to $\textbf{k}$  and $-\textbf{k}$ (or $ 2 \textbf{K} -\textbf{k} $ and $\textbf{k} - 2\textbf{K}$). 
In the chiral charge state, 
atoms of opposite spin at $\bf {K}$ and $\bf {K}$ are 
scattered to $\textbf{k}$  and $2 \textbf{K}-\textbf{k}$. 
Here the solid (dashed) arrows denote $\uparrow$ ($\downarrow$) and the red/blue colors differentiate 
between the two processes, which must be added symmetrically for every $\textbf{k}$.}
\label{fig:scattering} 
\end{figure}

\npsection{Emergence of chiral spin order.} 
We first illustrate a minimal model which supports a chiral superfluid ground state. We consider two component pseudo-spin-$1/2$ bosons in a spin-\textit{in}dependent optical lattice described by $H = H_0 + H_{\rm int}$, with 
\bea 
H_0 &=& \int \frac{d^d \tbf{k} }{(2\pi)^d} 
  \left( \epsilon(\tbf{k}) -\mu_\sigma \right) \phi_\sigma ^\dag (\tbf{k}) \phi_\sigma (\tbf{k}) \nn \\ 
H_{\rm int} &=& \frac{1}{2} \sum_{\sigma, \sigma', \tbf{x}, \tbf{x}'} U_{\sigma, \tbf{x}; \sigma', \tbf{x}'} 
    \phi_{\sigma\tbf{x}} ^\dag \phi_{\sigma' \tbf{x}'} ^\dag \phi_{\sigma' \tbf{x}'} \phi _{\sigma, \tbf{x}}, 
\eea 
where $\phi_{\sigma\tbf{x}} $ is the lattice annihilation operator with its 
Fourier transform 
$\phi_{\sigma} (\tbf{k} ) = \sum_\tbf{x} \phi_{\sigma\tbf{x}} 
e^{-i\tbf{k} \cdot \tbf{x}}$, $\epsilon (\tbf{k})$ 
is the energy dispersion, which is identical for both spin $\uparrow$ and $\downarrow$, $\mu_\sigma$ is the chemical potential and 
$U_{\sigma, \tbf{x}; \sigma', \tbf{x}'}$ is the density-density interaction. 
Our treatment in the following is valid for spatial dimensions $d=2$ or $3$. We study a situation where the single-particle dispersion $\epsilon(\textbf{k})$ possesses two minima, at generically incommensurate wave-vectors 
$\pm \textbf{K}$ related by TRS. Note that here TRS refers to an anti-unitary symmetry 
$T\phi_\sigma (\tbf{k}) T^{-1} = \phi_\sigma (-\tbf{k})$ under which spin is left unchanged, and 
the dispersion for such a system obeys $\epsilon (\tbf{k}) = \epsilon (-\tbf{k})$. 
This is because `spin' in our case simply refers to distinct hyperfine
states of an atom. Throughout, we will set 
$\epsilon (\pm \tbf{K}) =0 $ as the energy reference point. 
In the presence of translational symmetry, interactions preserve lattice momentum and
the coupling constant in momentum space is $U_{\sigma \sigma'} (\tbf{q}) =
\sum_\tbf{r} U_{\sigma, \tbf{x}; \sigma', \tbf{x}+\tbf{r} }e^{i\tbf{q} \cdot \tbf{r}}$. 

For weak interactions, the bosons condense at the two minima at $\pm {\bf K}$, and the condensate wave-function takes the form:
\bea
\varphi_{\sigma \tbf{r}} = \langle \phi_{\sigma, \tbf{r}} \rangle 
= \sqrt{\rho_{+,\sigma} } e^ {i\theta_{+,\sigma} } e^{i\tbf{K} \cdot \tbf{r}} 
+ \sqrt{\rho_{-,\sigma} } e^{i\theta_{-,\sigma} } e^{-i\tbf{K} \cdot \tbf{r} } 
\eea 
Here $\rho_{\pm,\sigma}$ refers to the density of each spin component at the $\pm \textbf{K}$ valleys, and $\theta_{\pm, \sigma}$ 
phases of the spin-component $\sigma$ at the two valleys. 
For single species bosons with short-ranged repulsion, the 
coexistence of $+\tbf{K}$ and $-\tbf{K}$ costs exchange interaction 
($U_{\sigma \sigma} (2\tbf{K}) >0$), so a
single-valley condensation associated with the spontaneous breaking of the valley symmetry is  energetically favorable. 
In the two component case we study, this  
exchange mechanism implies that 
$\rho_{+, \up} \rho_{-, \up} = \rho_{+, \down} \rho_{-, \down} = 0$, 
provided 
$ |U_{\up \down } (2\tbf{K}) | < \sqrt{U_{\up \up} (2\tbf{K}) U_{\down \down} (2\tbf{K})  } $, a condition
which is easily satisfied for weakly interacting repulsive spinor bosons. 
For contact interactions, this 
criterion reduces to the familiar criterion for macroscopic phase separation in real space~\cite{2008_Pethick_Smith}. 

Therefore at the mean-field level, each component condenses at a single momentum (either $+\tbf{K}$ or $-\tbf{K}$), 
yielding four degenerate choices for the condensate wavefunction $(\varphi_{\up, \tbf{r}} , \varphi_{\down, \tbf{r}} )$: 
$(++) \equiv (e^{i\tbf{K} \cdot \tbf{r}}, e^{i\tbf{K} \cdot \tbf{r}})$, 
$(+-)  \equiv (e^{i\tbf{K} \cdot \tbf{r}}, e^{-i\tbf{K} \cdot \tbf{r}} )$,
$(--) \equiv (e^{-i\tbf{K} \cdot \tbf{r}}, e^{-i\tbf{K} \cdot \tbf{r}}) $, and 
$(-+) \equiv (e^{-i\tbf{K} \cdot \tbf{r}}, e^{i\tbf{K} \cdot \tbf{r}} )$. 
The degeneracy of $(++)$ with $(--)$ [or $(+-)$ with $(-+)$ ] is guaranteed by TRS. 
However the degeneracy of $(++)$ with $(+-)$ is  
due to an accidental symmetry in the mean field energy, 
\bea 
E[\varphi_{\up \tbf{r}}, \varphi_{\down \tbf{r}} ^* ] 
= E[\varphi_{\up \tbf{r}}, \varphi_{\down \tbf{r}}  ],  
\eea 
with $E[\varphi_{\up \tbf{r}}, \varphi_{\down \tbf{r}} ]
 = H|_{\phi_{\sigma \tbf{r}}\to\varphi_{\sigma \tbf{r}} } $, 
resulting from the density-density nature of interactions which conserve the populations of each of the two spin components separately. 
In the $(++)$ or $(--)$ state, we have  chiral charge ($\chi_c$) order 
$
\int \frac{ d^d \tbf{k}} {(2\pi)^d}  \tbf{k} \langle \Phi ^\dag (\tbf{k}) \Phi (\tbf{k}) \rangle\neq 0,  
$ 
with $\Phi = (\phi_\up, \phi_\down) ^T$,  
while in the $(+-)$ or $(-+)$ state, we have chiral spin ($\chi_s$)  order 
$ 
\int \frac{d^d \tbf{k}} {(2\pi)^d} \tbf{k} \langle \Phi ^\dag  (\tbf{k}) \sigma_z \Phi (\tbf{k}) \rangle\neq 0. 
$  
In ultracold atom experiments, chiral spin and charge orders can be distinguished 
by using spin-resolved time-of-flight measurements~\cite{2012_Sengstock_NatPhys,2010_Demarco_NJPHYS}.

 
In the asymptotic weakly interacting limit, 
only the minimal momentum points $\pm \tbf{K}$ 
can be populated in the ground state, and the classical degeneracy is exact. We now investigate how quantum 
fluctuations lift this degeneracy through an ``order by disorder'' 
mechanism. To capture fluctuation effects, we start with a heuristic argument based on second order perturbation 
theory. The dominant inter-spin scattering processes which 
contribute to the energy correction for the $(++)$ and $(+-)$ 
(or equivalently the $(--)$ and $(-+)$) states at second order are shown in Fig.~\ref{fig:scattering} . 
Physically these processes correspond to annihilating  two condensate atoms in opposite 
spin states and creating two non-condensed atoms. For the chiral charge state, 
the two processes shown yield the same energy, and give rise to the first term in the right  hand 
side of Eq.~\eqref{eq:ediff}. By contrast, for the chiral spin state, the two processes produce different 
energy contributions given by the second and third terms in the right hand side of Eq.~\eqref{eq:ediff}. 

Treating these processes perturbatively, the resulting energy difference between the chiral spin and charge states 
$\Delta E ^{(2)} = E^{(2)} _ {\chi_c }- E^{(2)} _ {\chi_s }$ is readily obtained by integrating over momentum: 
\bea 
&& \Delta E^{(2)}  /N_s  
=  -\int \frac{d^d \tbf{k} } {(2\pi)^d} \rho_\up \rho_\down  
      \left\{  \frac{|U_{\up \down} (\tbf{k} - \tbf{K})|^2}
	  {\epsilon(\tbf{k}) + \epsilon(\tbf{Q}-\tbf{k}) } \right. \nn \\ 
&& \left. 	  -\frac{1}{2} \frac{|U_{\up \down} (\tbf{k} - \tbf{K})|^2 }
	  {\epsilon(\tbf{k}) + \epsilon(-\tbf{k}) }
	   -  \frac{1}{2} \frac{|U_{\up \down} (\tbf{K} - \tbf{k})|^2 }
	  {\epsilon(\tbf{Q} - \tbf{k}) + \epsilon(\tbf{k} - \tbf{Q} ) }
	    \right\}, 
\label{eq:ediff} 
\eea  
with $\tbf{Q} = 2\tbf{K}$, $N_s$ the total number of lattice sites, and 
the integral excludes the momentum $\tbf{k} = \pm\tbf{K}$ points. 
Using $\epsilon (\tbf{k}) = \epsilon(-\tbf{k})$, 
it follows from the relation $X^{-1} + Y^{-1} \ge {4}(X+Y)^{-1}$ (for positive numbers $X$ and $Y$) 
that 
$
 \Delta E^{(2)} > 0, $  
a remarkably {\it universal} result which is independent of the lattice geometry or details of the bandstructure.
The chiral spin superfluid state is generically selected, and this energetic selection rule is 
enforced by TRS. 

While the above argument is illuminating and captures the essential 
physics, there is a subtle issue in two dimensions because the integral in Eq.~\eqref{eq:ediff} 
is logarithmically divergent  (from the integral $\int d^2 \tbf{k} \frac{1}{ \tbf{k}^{2}} $). We thus need to go beyond the 
Rayleigh-Schr\"odinger type bare perturbation theory by performing a careful 
Bogoliubov theory analysis (akin to a renormalized Wigner-Brillouin type perturbation theory) 
in order to regularize the logarithmic divergence. In the 
renormalized theory (see Methods),  the bare dispersions in Eq.~\eqref{eq:ediff} are replaced by 
Bogoliubov energy dispersions, and the interaction $U_{\up\down}$ replaced by effective couplings
between the Bogoliubov quasiparticles. This cures the logarithmic divergence since Bogoliubov spectra appearing in the 
denominators are linear in momentum near the  condensate points. 
This improved analysis still yields the same robust universal result,
$
\Delta E^{(2)} >0,
$
generically favoring the chiral spin superfluid.

In three dimensions, the superfluid transition temperature of the chiral spin state is the usual Bose-Einstein condensate (BEC) 
temperature (see Methods). In two dimensions where phase fluctuations are strong, 
the superfluid transition temperature is determined 
by vortex proliferation associated with a Kosterlitz-Thouless (KT) transition at 
$T_c \approx \frac{\pi}{8} \sqrt{G} \rho_\sigma$, with 
$G$ being the curvature of the bandstructure at $\tbf{K}$ (see Methods) which determines the energy costs for phase twists. 
The chiral spin superfluid also breaks the discrete $Z_2$ symmetry, which is expected to be restored at a higher 
transition temperature~\cite{2014_Li_NatComm}, giving rise to a rich finite temperature phase diagram with an intermediate non-condensed 
chiral spin fluid phase separating the fully disordered and chiral spin superfluid phases.

Since the two spin components condense at opposite finite momenta in the chiral spin superfluid, the generic feature for this order is 
$\int d^d \tbf{k} \tbf{k} \left[ n_\up( \tbf{k}) - n_\down(\tbf{k})\right] \neq 0 $, which can be probed by spin-resolved time-of-flight measurements. 
Furthermore, when the Hamiltonian has TRS but no inversion symmetry, the chiral spin superfluid exhibits a 
spontaneous spin Hall effect, which is captured by the response to an applied linear potential (or a constant force ${\bf F}$), 
\bea 
\dot{\bf r}_s =		\pm {\bf F} \times \left( {\bf \Omega} (\tbf{K}) - {\bf \Omega } (- \tbf{K}) \right),  
\label{eq:rs}
\eea 
where ${ {\bf r}}_s$ is the vector connecting the charge centers of the two spin components, 
${\bf \Omega} (\tbf{k})$ is the Berry curvature~\cite{2010_Di_Chang_RMP}. 
In the interacting superfluid, this spin Hall effect only develops 
below the Ising transition associated with chiral order, and its sign fluctuates 
depending on which of the two Ising states the system picks, a spontaneously broken symmetry. Observation of the 
chiral spin superfluid with these novel properties would provide 
a direct demonstration of `quantum order by disorder', a quantum fluctuation effect beyond conventional mean field theories
used for BEC.

\begin{figure}[htp]
\includegraphics[angle=0,width=\linewidth]{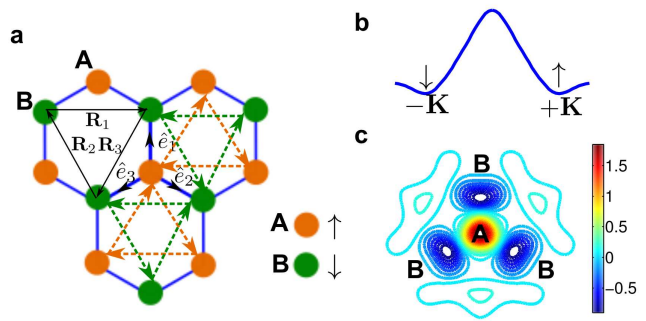}
\caption{The spin-dependent hexagonal lattice. {\bf a}, the lattice structure.  
Spin $\uparrow$ ($\downarrow$) bosons mainly live on $A$ ($B$) sites. 
{\bf b} , 
the condensate configuration for the chiral spin superfluid. 
Spin $\up$ and $\down$ condensing at $+\tbf{K}$ and $-\tbf{K}$ in the first excited band, 
corresponds to spin loop currents as  illustrated by dashed arrows in {\bf a}. 
{\bf c}, 
the contour plot of the excited band
Wannier function of spin $\up$, which has a peak at $A$ sites and dips at the $B$ sites 
(the opposite is true for the spin $\down$ Wannier function).} 
\label{fig:hexagonallattice} 
\end{figure}

\begin{figure}
\includegraphics[angle=0,width= \linewidth]{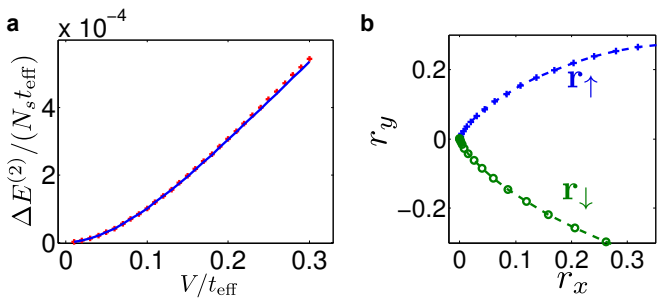}
\caption{ Emergence of chiral spin superfluid and the spin Hall effect. 
{\bf a}, the energy difference between chiral charge and spin superfluid states. 
The red `+' symbols are results calculated by solving the Bogoliubov spectra numerically and 
the blue solid line is from a combination with perturbation theory. 
See Methods section for the details of the tight binding Hamiltonian we solve. 
{\bf b}, the dynamics of the charge centers, ${\bf r}_{\up, \down}$, of the two spin components with 
a force $F$ applied in the $x$ direction. The two spins move oppositely in the transverse ($y$) direction, signifying a spin Hall effect. 
Here we choose $\Delta_D/t = 2$ and $F/t = 0.5$ ($t$ and $\Delta_D$ are nearest neighbor tunneling and Dirac gap, respectively, 
in the hexagonal lattice.) } 
\label{fig:Esplit} 
\end{figure}

\npsection{Spinor condensate at Dirac points.} 
We now consider a concrete model which exhibits the chiral superfluid ground state 
and the associated spin-Hall effect: bosonic atoms loaded at the massive 
Dirac points of a spin-{dependent} honeycomb lattice, shown in Fig.~\ref{fig:hexagonallattice}. 
Our choice is motivated by recent experiments, where two species of bosonic atoms 
have been loaded into the ground band of such a honeycomb lattice \cite{2012_Sengstock_NatPhys}. 

The  optical potential of the spin-dependent 
lattice is~\cite{2012_Sengstock_NatPhys} 
\bea 
&& \textstyle V_{\rm lattice} (\tbf{x}, m_F ) \nn \\ 
&=&\textstyle V_0 \left( \sum_j \cos (k \tbf{b}_j \cdot \tbf{x}) 
- m_F \alpha \sum_j \sin ( k \tbf{b}_j \cdot \tbf{x} ) \right),  
\label{eq:latpot}
\eea 
with $\tbf{b}_{j=0,1,2} = \left( - \sin ( \frac{2\pi}{3} j) , \cos (\frac{2\pi}{3} j)  \right)$.  
We set the lattice constant as the length unit. 
For $\alpha =0$, this potential has inversion symmetry, i.e., 
$V_{\rm lattice} (\tbf{x}, m_F ) =  V_{\rm lattice}(-\tbf{x}, m_F)$, and the realized hexagonal 
lattice has the bandstructure of graphene, two lowest bands touching at the Dirac points. 
With $\alpha \neq 0$, inversion symmetry is broken, and a gap opens at the Dirac points 
(akin to the Boron Nitride bandstructure~\cite{1994_Rubio_BN}). 
The first excited band has minima at two Dirac points (Fig.~\ref{fig:hexagonallattice}b), related by TRS. 
The model still has a combined spin-space inversion symmetry namely 
$V_{\rm lattice}(\tbf{x}, m_F ) = V_{\rm lattice}(-\tbf{x}, -m_F)$ which 
implies $\epsilon_{m_{F}}(\textbf{k}) = \epsilon_{-m_{F}}(-\textbf{k})$.
Using $m_F =1$ and $-1$ for pseudo-spin $\up$ and $\down$, the combined symmetry, 
together with TRS, guarantees that the two spin components  manifest the same 
energy dispersion $\epsilon (\tbf{k})$, so our general theory 
directly applies to this spin-dependent lattice Hamiltonian. We predict that 
the chiral spin superfluid state is the ground state for 
weakly interacting bosons loaded into the excited band of this lattice. 

\begin{figure}[htp]
\includegraphics[angle=0,width=\linewidth]{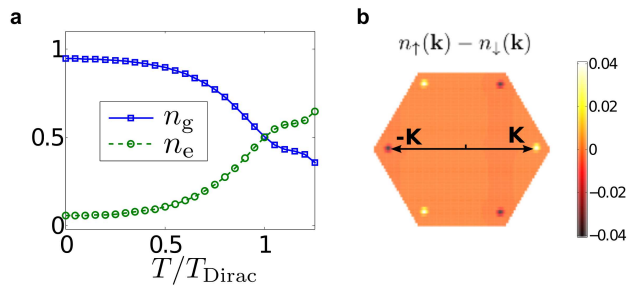}
\caption{ Simulation showing Bloch oscillations can be used to prepare the chiral spin superfluid state. 
{\bf a}, the occupation fractions of the ground and excited bands, $n_{\rm g}$ and $n_{\rm e}$, 
obtained by projecting the time-dependent wavefunction to the eigenmodes of $H_{\rm 2band}$ (see Methods). 
The time unit is $T _{\rm Dirac} \approx \frac{4 \pi \hbar}{3 \lambda }$. {\bf b}, the difference of 
momentum distributions of two spins $n_\up (\tbf{k}) - n_\down (\tbf{k})$ in the excited band 
at time $T_{\rm Dirac}$.
In our simulation 
we choose $\Delta_{\rm D} = 3 t$. The spread of momentum distribution over a finite momentum range is due to the harmonic 
trap included in the  simulation.   } 
\label{fig:BlochOscillate} 
\end{figure}

The energy splitting of chiral charge and spin states is shown in Fig.~\ref{fig:Esplit}a. 
The momentum distribution of the chiral spin superfluid state 
(Fig.~\ref{fig:BlochOscillate}b) has a similar pattern as the twisted superfluid reported 
in the experiment~\cite{2012_Sengstock_NatPhys},  but in our case, the 
condensates are at Dirac points rather than reciprocal lattice vectors.
Further because of the specialty of Bloch 
modes at Dirac points, the chiral spin superfluid actually has  staggered spin loop currents 
in real space (Fig.~\ref{fig:hexagonallattice}a), 
where  
spin and orbital motions are spontaneously coupled. 
In fermionic systems, spin loop current orderings are also found recently, 
albeit in a more delicate way~\cite{2008_Raghu_PRL,2014_Goto_arXiv}.  
Based on Eq.~\eqref{eq:rs}, the broken inversion symmetry implies that this chiral spin state exhibits a spin Hall effect 
due to the Berry curvature at a massive Dirac point, ${\bf \Omega}(-\tbf{K})  = - {\bf \Omega} (\tbf{K}) \neq 0$. 
(In condensed matter, this is also known as a ``valley Hall" effect~\cite{2007_Xiao_PRL}, 
but in our case valley and spin degrees of freedom are coupled in the chiral spin state, giving rise to a spin Hall effect.)  
In Fig.~\ref{fig:Esplit}b, we confirm this effect by direct numerical simulations.  
This chiral spin state and the consequent spin Hall effect can be prepared by Bloch Oscillations 
in a deterministic way in experiments (see Fig.~\ref{fig:BlochOscillate} and Methods).
 

\begin{figure}[htp]
\includegraphics[angle=0,width=\linewidth]{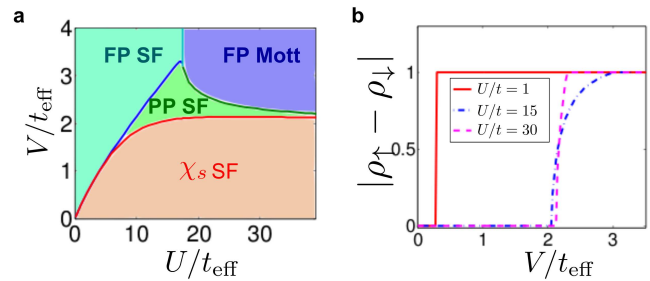}
\caption{ Phase diagram of spinor bosons in the spin dependent hexagonal lattice. {\bf a}, the phase 
diagram. The total density $\rho_\up + \rho_\down$ is fixed to be $1$ here. In the weakly interacting 
limit, the chiral spin ($\chi_s$) superfluid (SF) has a first order transition to the fully polarized 
(FP) superfluid state. A partially polarized (PP) SF state is stabilized in the intermediate interaction 
regime. A  FP Mott state shows in the stronger interaction regime. The phase boundaries of these polarized 
states meet at a quantum tricritical point. {\bf b}, the polarization as varying $V$ for different $U$s. } 
\label{fig:phasediag} 
\end{figure}

To study phase transitions from the weakly interacting chiral spin superfluid induced, 
we project into the second band, which is valid when the band gap $\Delta_D$ dominates over other energy scales 
such as 
effective tunneling $t_{\rm eff}$ in the second band, intra- and inter- species interactions $U$ and $V$ 
(see Methods). The Wannier orbitals for the spin $\uparrow$ and $\downarrow$ atoms in the resulting single 
band Hamiltonian (see Methods) are shown in Fig.~\ref{fig:hexagonallattice}c. We study the properties 
of this Hamiltonian using a Gutzwiller mean-field approach, which is 
known to predict the correct qualitative phase diagram in $d = 2, 3$. 

When $V$ is strong,  the chiral superfluid is unstable towards phase separation into fully 
polarized domains. In the weakly interacting limit, this transition is first order, and occurs at 
a critical interaction strength $V_c =  U/3$~\cite{2008_Pethick_Smith}. 
In the strong interaction limit $U\to \infty$, there is also a direct first order transition 
to a fully polarized state, yet at a different critical value $V_c = 2t_{\rm eff} $ (see Supplementary Information). 
In the intermediate regime, i.e., when $U$ and $V$ are not too large, correlation effects stabilize a 
partially polarized chiral spin superfluid state. 
The transitions out of this intermediate state are second order (Fig.~\ref{fig:phasediag}). 
With density fixed at 
$\rho_\up + \rho_\down = 1$, we find a novel quantum tricritical point at the crossing of the phase boundaries between these polarized phases. 

\npsection{Discussion} 

From our analysis, the chiral spin superfluid is a generic state for spinor Bose gases loaded into an energy band with 
double minima connected by time-reversal symmetry. This state thus not only exists in 
the hexagonal lattice~\cite{2012_Sengstock_NatPhys,2012_Esslinger_Nature}, 
but also in the $\pi$ flux triangular lattice~\cite{2013_Sengstock_NatPhys}, the 
shaken lattice~\cite{2013_Chin_NatPhys}, 
and other similar Bose systems. The generic feature for the chiral spin order, expected to emerge in all these 
setups is $\int d^d \tbf{k} \tbf{k} \left[ n_\up( \tbf{k}) - n_\down(\tbf{k})\right] \neq 0 $. 
In the shaken lattice setup~\cite{2013_Chin_NatPhys}, the chiral spin state would produce a time-of-flight signal similar to that
in spin-orbit coupled gases~\cite{2011_Lin_Spielman_Nature}, but with a {\it spontaneously chosen sign} for the spin-orbit coupling, 
which will vary from shot to shot. 
In optical lattices with broken inversion symmetry, this chiral spin superfluid supports a spontaneous spin Hall effect. The
required potential gradient to observe this transport phenomenon is different in spin-independent and spin-dependent lattices. 
For the former~\cite{2013_Chin_NatPhys,2012_Esslinger_Nature},
a spin-independent force (or non-magnetic potential gradient) is required, while 
for the latter~\cite{2012_Sengstock_NatPhys}, to accommodate spin-dependence of the lattice, 
a spin-dependent force $m_F {\bf F}$ (or a magnetic potential gradient) is necessary.

The idea that the chiral spin order is selected due to time-reversal invariance can be generalized 
to more general bands with multi-minima respecting crystalline symmetries, 
where nature of the momentum space magnetism is expected to be richer. Such multiple minimum bands are believed to 
occur in high spin systems such as Dysprosium or Erbium atom coupled to Raman fields~\cite{2013_Cui_PRAR}.  
In addition to the motivation from optical lattices~\cite{2012_Sengstock_NatPhys,2013_Chin_NatPhys,2013_Sengstock_NatPhys}, the proposed momentum space magnetism is potentially relevant 
to spontaneous vortex formation in BEC confined in ring geometries as well~\cite{2014_Snoke_arXiv}. 

\npsection{Methods}

\npsection{Regularization of the logarithmic divergence with Bogoliubov analysis. } 
To treat the logarithmic divergence problem of the bare perturbation theory (Eq.~\eqref{eq:ediff}) in two dimensions, 
a more careful analysis is performed by combining perturbation theory and Bogoliubov analysis, from which 
the energy difference between chiral spin and charge states is obtained to be (see Supplementary Information) 
\bea 
\label{eq:DeltaEbog} 
\textstyle 
&& \Delta E ^{(2)} /N_s = 
\textstyle -\frac{1}{2} {\rho_\up \rho_\down}
\int _\tbf{k}   g^2(\tbf{k})    \nn \\
&\times & \textstyle 
    \left\{  \frac{2 } {\varepsilon_\up (\tbf{k}, \tbf{Q} -\tbf{k}) 
			  + \varepsilon_\down(\tbf{k}, \tbf{Q} - \tbf{k}) } \right. \nn \\
	&& - \textstyle \frac{1}{\varepsilon_\up (\tbf{k}, \tbf{Q} -\tbf{k}) 
			  + \varepsilon_\down(-\tbf{Q} + \tbf{k}, -\tbf{k} )
			    + \Delta \epsilon  (\tbf{k},  \tbf{Q} - \tbf{k}) 
			    - \Delta \epsilon (-\tbf{Q} + \tbf{k}, -\tbf{k})  } \nn \\
&& -\left. \textstyle	
  \frac{ 1}{\varepsilon_\down (-\tbf{Q} +\tbf{k}, -\tbf{k}) 
		    + \varepsilon_\up (\tbf{k}, \tbf{Q} -\tbf{k}) 
		    + \Delta \epsilon (-\tbf{Q} + \tbf{k}, -\tbf{k}) 
		    - \Delta \epsilon (\tbf{k},  \tbf{Q} - \tbf{k} ) }
 \right\},  \nn 
\eea 
where 
$$
\varepsilon_\sigma ^2  (\tbf{k}_{1}, \tbf{k}_{2})  
	=  \overline{\epsilon} (\tbf{k}_{1}, \tbf{k}_{2} ) 
	    \left[ \overline{\epsilon} (\tbf{k}_{1}, \tbf{k}_{2}) + 2\rho_\sigma U_{\sigma \sigma} (\tbf{K} -\tbf{k})  \right], 
$$
$ \overline{\epsilon} ( \tbf{k}_{1} , \tbf{k}_{2} ) =
\left( \epsilon(\tbf{k}_{1}) + \epsilon (\tbf{k}_{2} ) \right) /{2}
$, 
and  
$\Delta \epsilon (\tbf{k}_{1} , \tbf{k}_{2}) = (\epsilon(\tbf{k}_{1}) - \epsilon(\tbf{k}_{2} ))/{2}$,  
and the effective couplings $g$ are given in Supplementary Information. 
With similar analysis as in the bare perturbation theory, 
we get  
$
\Delta E^{(2)} >0,
$ 
favoring the chiral spin superfluid generically.

\npsection{Finite temperature transitions.} 
In three dimensions at low temperature, the chiral spin superfluid state breaks two $U(1)$ symmetries (corresponding to two spins) and 
a $Z_2$ symmetry (${\bf k} \to -{\bf k}$). For the balanced case, with $\rho_\up = \rho_\down$, we expect three nearly coincident transitions 
(one Ising and two $U(1)$) near the three dimensional BEC transition temperature, while 
for the imbalanced case,  we expect a $U(1)$ transition for the minority spin at lower temperature, 
followed by nearly coincident transitions (Ising and $U(1)$ for the majority spin) at a higher temperature.
In two dimensions, the superfluidity transition temperature is determined by phase fluctuations. The fluctuations on 
top of chiral spin superfluid state are captured by introducing slowly varying fields $\varphi_{\sigma \tbf{r} } ^<$ as  
$ \varphi_{\uparrow \tbf{r}} = \varphi_{\uparrow \tbf{r}} ^< e^{i\tbf{K} \cdot \tbf{r}}$ 
and $\varphi_{\downarrow \tbf{r}} = \varphi_{\downarrow \tbf{r}} ^< e^{-i\tbf{K} \cdot \tbf{r}} $. The energy cost of these fluctuations 
is 
\bea 
\Delta E = \int d^2 \tbf{x} \sum_\sigma
 \left\{ 	\frac{1}{2} Z_{i j} \partial_{x_i} \varphi_{\sigma,\tbf{x}} ^{<*} \partial_{x_j} \varphi_{\sigma,\tbf{x} }^{<}   \right\},  \nn 
\eea 
with $Z_{i j} =  \partial_{k_i} \partial_{k_j} \epsilon (\tbf{k}) |_{\tbf{k} \to \tbf{K}} $. Transforming to the coordinate frame 
with $Z_{i j}$ being diagonal and replacing $\varphi_{\sigma\tbf{x}} ^<$ by $\sqrt{\rho_\sigma} e^{i\theta_\sigma} $, $\Delta E$ is rewritten as  
$
\Delta E  = \frac{1}{2} \int d^2 \tbf{x} \sum_\sigma \rho_\sigma 
  \left\{ \lambda_1 (\partial_{x_1} \theta_\sigma )^2 + \lambda_2 (\partial_{x_2} \theta_\sigma) ^2 \right\}, 
$ 
where $\lambda_{1,2}$ are  eigenvalues of $[Z]$. 
The KT transition temperature is then estimated to be 
$
T_c \approx \frac{\pi}{8} \sqrt{G} \rho_\sigma, 
$ 
with $G$ the Gaussian curvature of the bandstructure at $\tbf{K}$, which is $\lambda_1 \lambda_2$. 
For the symmetric case with $\rho_\up = \rho_\down$, we have one single KT transition temperature, while for 
$\rho_\up \neq \rho_\down$ there are two separate KT transitions at two distinct temperatures. The Ising transition
associated with the chiral order is expected to occur slightly above higher superfluid transition, as observed in other
studies of chiral superfluids \cite{2014_Li_NatComm}. 
In principle, a chiral spin state which has chiral spin order but no superfluidity could occur in a temperature window above 
superfluid transitions~\cite{2014_Li_NatComm}; the exploration of such a remarkable bosonic chiral spin fluid is left for future studies.

\npsection{Experimental preparation of the chiral spin state.} 
Here we propose a deterministic way to prepare the chiral superfluid state with Bloch oscillations. 
We could start with the lowest band condensate in the  lattice potential 
$V_{\rm lattice} (\tbf{x}, m_F) |_{\alpha\to 0} $ for which the two lowest bands touch at Dirac points. 
Applying a magnetic gradient potential $-m_F \lambda x$, spin $\up$ and $\down$ components will move 
towards the Dirac points at $\tbf{K}$ and $-\tbf{K}$, respectively. At time $T _{\rm Dirac} \approx \frac{4 \pi \hbar}{3 \lambda}$, 
the components reach the respective Dirac points, after which we quickly ramp on the spin dependent potential 
(the term proportional to $\alpha$ in Eq.~\eqref{eq:latpot}), to make the Dirac points massive with a band 
gap $\Delta_{\rm D}$. With $\Delta_{\rm D}$ much larger than the bandwidth and  interactions, the inter-band dynamics 
will be greatly suppressed. The meta-stable state in the excited band is given by an effective 
single-band Hamiltonian, described in the next paragraph. To demonstrate the efficiency of the proposed procedure, we simulate 
the Bloch oscillations by taking 
a two-band tight binding model of free bosons, 
\bea
&& H_{\rm 2band} = \textstyle -t \sum_{<\tbf{r}, \tbf{r}'>}  \left(\phi_{A\sigma, \tbf{r} } ^\dag \phi_{B \sigma, \tbf{r}} +h.c. \right) \nn  \\ 
 &&    +  \frac{1}{2} m_F\Delta_{\rm D} \sum_\tbf{r} 
    \left(\phi_{A\sigma, \tbf{r}} ^\dag \phi_{A\sigma, \tbf{r}}  - \phi_{B\sigma, \tbf{r}+\hat{e}_1} ^\dag   \phi_{B\sigma, \tbf{r}+\hat{e}_1} \right),  \nn
\label{eq:2bandHam} 
\eea
and the  magnetic gradient potential is modeled as 
\bea 
H_{\rm linear} = -\lambda m_F \sum_\tbf{r}  r_x \left(\phi_{A\sigma, \tbf{r}} ^\dag \phi_{A\sigma, \tbf{r}}
					    + \phi_{B\sigma, \tbf{r}+\hat{e}_1} ^\dag   \phi_{B\sigma, \tbf{r}+\hat{e}_1} \right), \nn 
\eea 
where $A$ and $B$ label two sublattices as shown in Fig.~\ref{fig:hexagonallattice}. 
We find that the occupation fraction of the excited band could easily reach $50\%$ (Fig.~\ref{fig:BlochOscillate}).

\npsection{Calculation of the phase diagram.} 
To obtain the phase diagram of the meta-stable states in the second band of the hexagonal lattice, 
we construct an effective single band tight binding model, 
\bea 
H_0 &=& \sum_{\tbf{r}, j} t_{\rm eff}  \left[  \phi_{\uparrow, \tbf{r} } ^\dag \phi_{\uparrow, \tbf{r} + \tbf{R}_j}   
				+ \phi_{\downarrow, \tbf{r} + \hat{e}_1 }  ^\dag  \phi_{\downarrow, \tbf{r} + \hat{e}_1 + \tbf{R}_j } 
					+  h.c. \right], \nn \\
H_{\rm int} &=&  \frac{U}{2} \sum_{\tbf{r}} 
	\left\{ n_{\uparrow, \tbf{r}} (n_{\uparrow, \tbf{r}}-1) 
		+ n_{\downarrow, \tbf{r} + \hat{e}_1} (n_{\downarrow, \tbf{r} + \hat{e}_1} -1) 
    \right\} \nn \\
		&+& V \sum_{\tbf{r}, j} n_{\uparrow, \tbf{r}}  n_{\downarrow, \tbf{r} + \hat{e}_j}. \nn 
\eea 
Here $\phi_{\sigma, \tbf{r}}$ is the annihilation operator for the Wannier functions peaked at position $\tbf{r}$ 
(Fig.~\ref{fig:hexagonallattice}), and each spin species sees a triangular lattice. 
This Bose-Hubbard model describes bosons loaded into the second band of the hexagonal lattice. 
In this lattice setup, the Wannier 
functions of $\up$ and $\down$ components are peaked at two nearby sites rather than on 
the same one, which makes the ratio of interactions $V/U$ easily tunable. For example this 
ratio can be decreased by increasing the lattice depth or the spin dependence parameter $\alpha$. 
The energy dispersion from the tight binding model is 
$
\epsilon (\tbf{k}) = 2 t \sum_j \cos ( \tbf{k} \cdot \tbf{R}_j),  
$
which has band minima at $\pm \tbf{K} = (\pm \frac{4\pi}{3},0)$. 
For weak interactions, the energy difference between the chiral charge 
and spin states computed from this model is shown in Fig.~\ref{fig:Esplit}. 

\npsection{Acknowledgements} 

We thank W. Vincent Liu, Di Xiao, Kai Sun,  R. G. Hulet, and D. Snoke for helpful discussions.  
This work is supported by JQI-NSF-PFC and ARO-Atomtronics-MURI(X.L., S.N. and S.D.S.). 
A.P. acknowledges support from NSERC of Canada.

\npsection{Author contributions} 

X.L. conceived the theoretical ideas and performed calculations with insightful 
inputs from S.N., A.P., S.D.S. All authors worked on theoretical analysis 
and contributed to the manuscript preparation.   

\bibliographystyle{naturemag} 
\bibliography{chispinsf}

\begin{thebibliography}{10}
\expandafter\ifx\csname url\endcsname\relax
  \def\url#1{\texttt{#1}}\fi
\expandafter\ifx\csname urlprefix\endcsname\relax\def\urlprefix{URL }\fi
\providecommand{\bibinfo}[2]{#2}
\providecommand{\eprint}[2][]{\url{#2}}

\bibitem{2004_DasSarma_RMP}
\bibinfo{author}{\ifmmode \check{Z}\else \v{Z}\fi{}uti\ifmmode~\acute{c}\else
  \'{c}\fi{}, I.}, \bibinfo{author}{Fabian, J.} \& \bibinfo{author}{Das~Sarma,
  S.}
\newblock \bibinfo{title}{Spintronics: Fundamentals and applications}.
\newblock \emph{\bibinfo{journal}{Rev. Mod. Phys.}}
  \textbf{\bibinfo{volume}{76}}, \bibinfo{pages}{323--410}
  (\bibinfo{year}{2004}).

\bibitem{2007_Xiao_PRL}
\bibinfo{author}{Xiao, D.}, \bibinfo{author}{Yao, W.} \& \bibinfo{author}{Niu,
  Q.}
\newblock \bibinfo{title}{Valley-contrasting physics in graphene: Magnetic
  moment and topological transport}.
\newblock \emph{\bibinfo{journal}{Phys. Rev. Lett.}}
  \textbf{\bibinfo{volume}{99}}, \bibinfo{pages}{236809}
  (\bibinfo{year}{2007}).

\bibitem{2002_StamperKurn_PRL}
\bibinfo{author}{Higbie, J.} \& \bibinfo{author}{Stamper-Kurn, D.~M.}
\newblock \bibinfo{title}{Periodically dressed bose-einstein condensate: A
  superfluid with an anisotropic and variable critical velocity}.
\newblock \emph{\bibinfo{journal}{Phys. Rev. Lett.}}
  \textbf{\bibinfo{volume}{88}}, \bibinfo{pages}{090401}
  (\bibinfo{year}{2002}).

\bibitem{2009_Lin_Nat}
\bibinfo{author}{Lin, Y.-J.}, \bibinfo{author}{Compton, R.~L.},
  \bibinfo{author}{Jimenez-Garcia, K.}, \bibinfo{author}{Porto, J.~V.} \&
  \bibinfo{author}{Spielman, I.~B.}
\newblock \bibinfo{title}{Synthetic magnetic fields for ultracold neutral
  atoms}.
\newblock \emph{\bibinfo{journal}{Nature}} \textbf{\bibinfo{volume}{462}},
  \bibinfo{pages}{628--632} (\bibinfo{year}{2009}).

\bibitem{2011_Lin_Spielman_Nature}
\bibinfo{author}{Lin, Y.-J.}, \bibinfo{author}{Jimenez-Garcia, K.} \&
  \bibinfo{author}{Spielman, I.~B.}
\newblock \bibinfo{title}{Spin-orbit-coupled bose-einstein condensates}.
\newblock \emph{\bibinfo{journal}{Nature}} \textbf{\bibinfo{volume}{471}},
  \bibinfo{pages}{83--86} (\bibinfo{year}{2011}).

\bibitem{2011_Hemmerich_NatPhys}
\bibinfo{author}{Wirth, G.}, \bibinfo{author}{Olschlager, M.} \&
  \bibinfo{author}{Hemmerich, A.}
\newblock \bibinfo{title}{Evidence for orbital superfluidity in the p-band of a
  bipartite optical square lattice}.
\newblock \emph{\bibinfo{journal}{Nat Phys}} \textbf{\bibinfo{volume}{7}},
  \bibinfo{pages}{147--153} (\bibinfo{year}{2011}).

\bibitem{2012_Chen_Pan_PRL}
\bibinfo{author}{Zhang, J.-Y.} \emph{et~al.}
\newblock \bibinfo{title}{Collective dipole oscillations of a spin-orbit
  coupled bose-einstein condensate}.
\newblock \emph{\bibinfo{journal}{Phys. Rev. Lett.}}
  \textbf{\bibinfo{volume}{109}}, \bibinfo{pages}{115301}
  (\bibinfo{year}{2012}).

\bibitem{2012_Zhang_PRL}
\bibinfo{author}{Wang, P.} \emph{et~al.}
\newblock \bibinfo{title}{Spin-orbit coupled degenerate fermi gases}.
\newblock \emph{\bibinfo{journal}{Phys. Rev. Lett.}}
  \textbf{\bibinfo{volume}{109}}, \bibinfo{pages}{095301}
  (\bibinfo{year}{2012}).

\bibitem{2012_Zwierlein_PRL}
\bibinfo{author}{Cheuk, L.~W.} \emph{et~al.}
\newblock \bibinfo{title}{Spin-injection spectroscopy of a spin-orbit coupled
  fermi gas}.
\newblock \emph{\bibinfo{journal}{Phys. Rev. Lett.}}
  \textbf{\bibinfo{volume}{109}}, \bibinfo{pages}{095302}
  (\bibinfo{year}{2012}).

\bibitem{2012_Sengstock_NatPhys}
\bibinfo{author}{Soltan-Panahi, P.}, \bibinfo{author}{Luhmann, D.-S.},
  \bibinfo{author}{Struck, J.}, \bibinfo{author}{Windpassinger, P.} \&
  \bibinfo{author}{Sengstock, K.}
\newblock \bibinfo{title}{Quantum phase transition to unconventional
  multi-orbital superfluidity in optical lattices}.
\newblock \emph{\bibinfo{journal}{Nat Phys}} \textbf{\bibinfo{volume}{8}},
  \bibinfo{pages}{71--75} (\bibinfo{year}{2012}).

\bibitem{2012_Esslinger_Nature}
\bibinfo{author}{Tarruell, L.}, \bibinfo{author}{Greif, D.},
  \bibinfo{author}{Uehlinger, T.}, \bibinfo{author}{Jotzu, G.} \&
  \bibinfo{author}{Esslinger, T.}
\newblock \bibinfo{title}{Creating, moving and merging dirac points with a
  fermi gas in a tunable honeycomb lattice}.
\newblock \emph{\bibinfo{journal}{Nature}} \textbf{\bibinfo{volume}{483}},
  \bibinfo{pages}{302--305} (\bibinfo{year}{2012}).

\bibitem{2013_Chin_NatPhys}
\bibinfo{author}{Parker, C.~V.}, \bibinfo{author}{Ha, L.-C.} \&
  \bibinfo{author}{Chin, C.}
\newblock \bibinfo{title}{Direct observation of effective ferromagnetic domains
  of cold atoms in a shaken optical lattice}.
\newblock \emph{\bibinfo{journal}{Nat Phys}} \textbf{\bibinfo{volume}{9}},
  \bibinfo{pages}{769--774} (\bibinfo{year}{2013}).

\bibitem{2013_Sengstock_NatPhys}
\bibinfo{author}{Struck, J.} \emph{et~al.}
\newblock \bibinfo{title}{Engineering ising-xy spin-models in a triangular
  lattice using tunable artificial gauge fields}.
\newblock \emph{\bibinfo{journal}{Nat Phys}} \textbf{\bibinfo{volume}{9}},
  \bibinfo{pages}{738--743} (\bibinfo{year}{2013}).

\bibitem{2013_Hemmerich_NJP}
\bibinfo{author}{{{\"O}lschl{\"a}ger}, M.} \emph{et~al.}
\newblock \bibinfo{title}{Interaction-induced chiral $p_x \pm ip_y$ superfluid
  order of bosons in an optical lattice}.
\newblock \emph{\bibinfo{journal}{New Journal of Physics}}
  \textbf{\bibinfo{volume}{15}}, \bibinfo{pages}{083041}
  (\bibinfo{year}{2013}).

\bibitem{2013_Bloch_PRL}
\bibinfo{author}{Aidelsburger, M.} \emph{et~al.}
\newblock \bibinfo{title}{Realization of the hofstadter hamiltonian with
  ultracold atoms in optical lattices}.
\newblock \emph{\bibinfo{journal}{Phys. Rev. Lett.}}
  \textbf{\bibinfo{volume}{111}}, \bibinfo{pages}{185301}
  (\bibinfo{year}{2013}).

\bibitem{2013_Wolfgang_PRL}
\bibinfo{author}{Miyake, H.}, \bibinfo{author}{Siviloglou, G.~A.},
  \bibinfo{author}{Kennedy, C.~J.}, \bibinfo{author}{Burton, W.~C.} \&
  \bibinfo{author}{Ketterle, W.}
\newblock \bibinfo{title}{Realizing the harper hamiltonian with laser-assisted
  tunneling in optical lattices}.
\newblock \emph{\bibinfo{journal}{Phys. Rev. Lett.}}
  \textbf{\bibinfo{volume}{111}}, \bibinfo{pages}{185302}
  (\bibinfo{year}{2013}).

\bibitem{2014_Bloch_arXiv}
\bibinfo{author}{{Atala}, M.} \emph{et~al.}
\newblock \bibinfo{title}{{Observation of the Meissner effect with ultracold
  atoms in bosonic ladders}}.
\newblock \emph{\bibinfo{journal}{ArXiv e-prints}}  (\bibinfo{year}{2014}).
\newblock \eprint{1402.0819}.

\bibitem{2007_Esslinger_PRL}
\bibinfo{author}{Strohmaier, N.} \emph{et~al.}
\newblock \bibinfo{title}{Interaction-controlled transport of an ultracold
  fermi gas}.
\newblock \emph{\bibinfo{journal}{Phys. Rev. Lett.}}
  \textbf{\bibinfo{volume}{99}}, \bibinfo{pages}{220601}
  (\bibinfo{year}{2007}).

\bibitem{2013_Chin_arXiv}
\bibinfo{author}{{Hazlett}, E.~L.}, \bibinfo{author}{{Ha}, L.-C.} \&
  \bibinfo{author}{{Chin}, C.}
\newblock \bibinfo{title}{{Anomalous thermoelectric transport in
  two-dimensional Bose gas}}.
\newblock \emph{\bibinfo{journal}{ArXiv e-prints}}  (\bibinfo{year}{2013}).
\newblock \eprint{1306.4018}.

\bibitem{2013_Spielman_Nature}
\bibinfo{author}{Beeler, M.~C.} \emph{et~al.}
\newblock \bibinfo{title}{The spin hall effect in a quantum gas}.
\newblock \emph{\bibinfo{journal}{Nature}} \textbf{\bibinfo{volume}{498}},
  \bibinfo{pages}{201--204} (\bibinfo{year}{2013}).

\bibitem{2014_Esslinger_arXiv}
\bibinfo{author}{{Krinner}, S.}, \bibinfo{author}{{Stadler}, D.},
  \bibinfo{author}{{Husmann}, D.}, \bibinfo{author}{{Brantut}, J.-P.} \&
  \bibinfo{author}{{Esslinger}, T.}
\newblock \bibinfo{title}{{Observation of Quantized Conductance in Neutral
  Matter}}.
\newblock \emph{\bibinfo{journal}{ArXiv e-prints}}  (\bibinfo{year}{2014}).
\newblock \eprint{1404.6400}.

\bibitem{2013_Galitski_Nat}
\bibinfo{author}{Galitski, V.} \& \bibinfo{author}{Spielman, I.~B.}
\newblock \bibinfo{title}{Spin-orbit coupling in quantum gases}.
\newblock \emph{\bibinfo{journal}{Nature}} \textbf{\bibinfo{volume}{494}},
  \bibinfo{pages}{49--54} (\bibinfo{year}{2013}).

\bibitem{2011_Liu_NatPhys}
\bibinfo{author}{Lewenstein, M.} \& \bibinfo{author}{Liu, W.~V.}
\newblock \bibinfo{title}{Optical lattices: Orbital dance}.
\newblock \emph{\bibinfo{journal}{Nat Phys}} \textbf{\bibinfo{volume}{7}},
  \bibinfo{pages}{101--103} (\bibinfo{year}{2011}).

\bibitem{1982_Tsui_PRL}
\bibinfo{author}{Tsui, D.~C.}, \bibinfo{author}{Stormer, H.~L.} \&
  \bibinfo{author}{Gossard, A.~C.}
\newblock \bibinfo{title}{Two-dimensional magnetotransport in the extreme
  quantum limit}.
\newblock \emph{\bibinfo{journal}{Phys. Rev. Lett.}}
  \textbf{\bibinfo{volume}{48}}, \bibinfo{pages}{1559--1562}
  (\bibinfo{year}{1982}).

\bibitem{2013_Chen_PRL}
\bibinfo{author}{Chen, G.} \& \bibinfo{author}{Balents, L.}
\newblock \bibinfo{title}{Ferromagnetism in itinerant two-dimensional $t_{2g}$
  systems}.
\newblock \emph{\bibinfo{journal}{Phys. Rev. Lett.}}
  \textbf{\bibinfo{volume}{110}}, \bibinfo{pages}{206401}
  (\bibinfo{year}{2013}).

\bibitem{2013_Wu_arXiv}
\bibinfo{author}{{Li}, Y.}, \bibinfo{author}{{Lieb}, E.~H.} \&
  \bibinfo{author}{{Wu}, C.}
\newblock \bibinfo{title}{{Exact Results on Itinerant Ferromagnetism in
  Multi-orbital Systems on Square and Cubic Lattices}}.
\newblock \emph{\bibinfo{journal}{ArXiv e-prints}}  (\bibinfo{year}{2013}).
\newblock \eprint{1310.4391}.

\bibitem{2010_Balents_Nature}
\bibinfo{author}{Balents, L.}
\newblock \bibinfo{title}{Spin liquids in frustrated magnets}.
\newblock \emph{\bibinfo{journal}{Nature}} \textbf{\bibinfo{volume}{464}},
  \bibinfo{pages}{199--208} (\bibinfo{year}{2010}).

\bibitem{1980_Villain_JPF}
\bibinfo{author}{{Villain, J.}}, \bibinfo{author}{{Bidaux, R.}},
  \bibinfo{author}{{Carton, J.-P.}} \& \bibinfo{author}{{Conte, R.}}
\newblock \bibinfo{title}{Order as an effect of disorder}.
\newblock \emph{\bibinfo{journal}{J. Phys. France}}
  \textbf{\bibinfo{volume}{41}}, \bibinfo{pages}{1263--1272}
  (\bibinfo{year}{1980}).

\bibitem{1989_Henley_PRL}
\bibinfo{author}{Henley, C.~L.}
\newblock \bibinfo{title}{Ordering due to disorder in a frustrated vector
  antiferromagnet}.
\newblock \emph{\bibinfo{journal}{Phys. Rev. Lett.}}
  \textbf{\bibinfo{volume}{62}}, \bibinfo{pages}{2056--2059}
  (\bibinfo{year}{1989}).

\bibitem{2012_Paramekanti_PRA}
\bibinfo{author}{Dhar, A.} \emph{et~al.}
\newblock \bibinfo{title}{Bose-hubbard model in a strong effective magnetic
  field: Emergence of a chiral mott insulator ground state}.
\newblock \emph{\bibinfo{journal}{Phys. Rev. A}} \textbf{\bibinfo{volume}{85}},
  \bibinfo{pages}{041602} (\bibinfo{year}{2012}).

\bibitem{2012_You_PRL}
\bibinfo{author}{You, Y.-Z.}, \bibinfo{author}{Chen, Z.}, \bibinfo{author}{Sun,
  X.-Q.} \& \bibinfo{author}{Zhai, H.}
\newblock \bibinfo{title}{Superfluidity of bosons in kagome lattices with
  frustration}.
\newblock \emph{\bibinfo{journal}{Phys. Rev. Lett.}}
  \textbf{\bibinfo{volume}{109}}, \bibinfo{pages}{265302}
  (\bibinfo{year}{2012}).

\bibitem{2013_Paramekanti_PRB}
\bibinfo{author}{Dhar, A.} \emph{et~al.}
\newblock \bibinfo{title}{Chiral mott insulator with staggered loop currents in
  the fully frustrated bose-hubbard model}.
\newblock \emph{\bibinfo{journal}{Phys. Rev. B}} \textbf{\bibinfo{volume}{87}},
  \bibinfo{pages}{174501} (\bibinfo{year}{2013}).

\bibitem{2014_Zaletel_PRB}
\bibinfo{author}{Zaletel, M.~P.}, \bibinfo{author}{Parameswaran, S.~A.},
  \bibinfo{author}{R\"uegg, A.} \& \bibinfo{author}{Altman, E.}
\newblock \bibinfo{title}{Chiral bosonic mott insulator on the frustrated
  triangular lattice}.
\newblock \emph{\bibinfo{journal}{Phys. Rev. B}} \textbf{\bibinfo{volume}{89}},
  \bibinfo{pages}{155142} (\bibinfo{year}{2014}).

\bibitem{2008_Pethick_Smith}
\bibinfo{author}{Pethick, C.} \& \bibinfo{author}{Smith, H.}
\newblock \emph{\bibinfo{title}{Bose-Einstein Condensation in dilute gases}}
  (\bibinfo{publisher}{cambridge university press}, \bibinfo{year}{2008}).

\bibitem{2010_Demarco_NJPHYS}
\bibinfo{author}{McKay, D.} \& \bibinfo{author}{DeMarco, B.}
\newblock \bibinfo{title}{Thermometry with spin-dependent lattices}.
\newblock \emph{\bibinfo{journal}{New Journal of Physics}}
  \textbf{\bibinfo{volume}{12}}, \bibinfo{pages}{055013}
  (\bibinfo{year}{2010}).

\bibitem{2014_Li_NatComm}
\bibinfo{author}{Li, X.}, \bibinfo{author}{Paramekanti, A.},
  \bibinfo{author}{Hemmerich, A.} \& \bibinfo{author}{Liu, W.~V.}
\newblock \bibinfo{title}{Proposed formation and dynamical signature of a
  chiral bose liquid in an optical lattice}.
\newblock \emph{\bibinfo{journal}{Nat Commun}} \textbf{\bibinfo{volume}{5}},
  \bibinfo{pages}{--} (\bibinfo{year}{2014}).

\bibitem{2010_Di_Chang_RMP}
\bibinfo{author}{Xiao, D.}, \bibinfo{author}{Chang, M.-C.} \&
  \bibinfo{author}{Niu, Q.}
\newblock \bibinfo{title}{Berry phase effects on electronic properties}.
\newblock \emph{\bibinfo{journal}{Rev. Mod. Phys.}}
  \textbf{\bibinfo{volume}{82}}, \bibinfo{pages}{1959--2007}
  (\bibinfo{year}{2010}).

\bibitem{1994_Rubio_BN}
\bibinfo{author}{Rubio, A.}, \bibinfo{author}{Corkill, J.~L.} \&
  \bibinfo{author}{Cohen, M.~L.}
\newblock \bibinfo{title}{Theory of graphitic boron nitride nanotubes}.
\newblock \emph{\bibinfo{journal}{Phys. Rev. B}} \textbf{\bibinfo{volume}{49}},
  \bibinfo{pages}{5081--5084} (\bibinfo{year}{1994}).

\bibitem{2008_Raghu_PRL}
\bibinfo{author}{Raghu, S.}, \bibinfo{author}{Qi, X.-L.},
  \bibinfo{author}{Honerkamp, C.} \& \bibinfo{author}{Zhang, S.-C.}
\newblock \bibinfo{title}{Topological mott insulators}.
\newblock \emph{\bibinfo{journal}{Phys. Rev. Lett.}}
  \textbf{\bibinfo{volume}{100}}, \bibinfo{pages}{156401}
  (\bibinfo{year}{2008}).

\bibitem{2014_Goto_arXiv}
\bibinfo{author}{{Goto}, S.}, \bibinfo{author}{{Masuda}, K.} \&
  \bibinfo{author}{{Kurihara}, S.}
\newblock \bibinfo{title}{{Spontaneous loop-spin current with topological
  characters in the Hubbard model}}.
\newblock \emph{\bibinfo{journal}{ArXiv e-prints}}  (\bibinfo{year}{2014}).
\newblock \eprint{1405.3808}.

\bibitem{2013_Cui_PRAR}
\bibinfo{author}{Cui, X.}, \bibinfo{author}{Lian, B.}, \bibinfo{author}{Ho,
  T.-L.}, \bibinfo{author}{Lev, B.~L.} \& \bibinfo{author}{Zhai, H.}
\newblock \bibinfo{title}{Synthetic gauge field with highly magnetic lanthanide
  atoms}.
\newblock \emph{\bibinfo{journal}{Phys. Rev. A}} \textbf{\bibinfo{volume}{88}},
  \bibinfo{pages}{011601} (\bibinfo{year}{2013}).

\bibitem{2014_Snoke_arXiv}
\bibinfo{author}{{Liu}, G.}, \bibinfo{author}{{Snoke}, D.~W.},
  \bibinfo{author}{{Daley}, A.}, \bibinfo{author}{{Pfeiffer}, L.} \&
  \bibinfo{author}{{West}, K.}
\newblock \bibinfo{title}{{Spin-Flipping Half Vortex in a Macroscopic Polariton
  Spinor Ring Condensate}}.
\newblock \emph{\bibinfo{journal}{ArXiv e-prints}}  (\bibinfo{year}{2014}).
\newblock \eprint{1402.4339}.

\bibitem{2013_Stamper_Kurn_RMP}
\bibinfo{author}{Stamper-Kurn, D.~M.} \& \bibinfo{author}{Ueda, M.}
\newblock \bibinfo{title}{Spinor bose gases: Symmetries, magnetism, and quantum
  dynamics}.
\newblock \emph{\bibinfo{journal}{Rev. Mod. Phys.}}
  \textbf{\bibinfo{volume}{85}}, \bibinfo{pages}{1191--1244}
  (\bibinfo{year}{2013}).

\end{thebibliography}

\onecolumngrid

\newpage 

\renewcommand{\thesection}{S-\arabic{section}}
\renewcommand{\theequation}{S\arabic{equation}}
\setcounter{equation}{0}  
\renewcommand{\thefigure}{S\arabic{figure}}
\setcounter{figure}{0}  

\section*{\Large\bf Supplementary Information}
\begin{center}
{\large \bf \papertitle} 
\end{center}

\section{Details of the Bogoliubov analysis} 
\label{sec:BogDetails}

In this section, we give the details of our Bogoliubov analysis. The ground state energy corrections are explicitly calculated for 
the $(++)$ and $(+-)$ states. Symmetry guarantees the same result for  the $(--)$ and $(-+)$ states.  Fluctuations 
are included as 
\bea 
\phi_{\uparrow \tbf{r}} &=& \sqrt{\rho_\uparrow} e^{ i\tbf{K} \cdot \tbf{r}} 
 + \int _\tbf{k}  \phi_\uparrow(\tbf{k}) e^{i\tbf{k} \cdot \tbf{r}}  \nn \\ 
 \phi_{\downarrow \tbf{r}} &=& \sqrt{\rho_\downarrow} e^{ip\tbf{K} \cdot \tbf{r}} 
 + \int _\tbf{k}   \phi_\downarrow (p\tbf{k}) e^{ip\tbf{k} \cdot \tbf{r}}, 
\eea 
with $p=$ $+$ and $-$ for chiral charge and spin states, respectively. 
The effective Bogoliubov Hamiltonian controlling these fluctuations is 
\bea 
H _{\rm eff} = \frac{1}{2} \int  _\tbf{k}  \Psi ^\dag  (\tbf{k}) {\cal H} (\tbf{k} ) \Psi   (\tbf{k}) + const, 
\label{eq:SHeff}
\eea 
where $\Psi (\tbf{k}) = 
[ \phi_\uparrow (\tbf{k}_{1\uparrow} ), \phi_\uparrow ^\dag (\tbf{k}_{2\uparrow} ), 
\phi_\downarrow (\tbf{k}_{1\down} ), \phi_\downarrow ^\dag  (\tbf{k}_{2\down}) ]^T $ and 
\be 
{\cal H } ( \tbf{k}) = \left [ 
			\begin{array}{cc} 
			M_{\uparrow \uparrow} (\tbf{k}_{1\up}, \tbf{k}_{2\up} ) & M_{\uparrow \downarrow} (\tbf{k}) \\ 
			M_{\downarrow \uparrow} (\tbf{k}) & M_{\downarrow \downarrow} (\tbf{k}_{1\down}, \tbf{k}_{2\down} )  
			\end{array} 
			\right] , 
\label{eq:Hkmat} 
\ee 
with $(\tbf{k}_{1\up}, \tbf{k}_{2\up}, \tbf{k}_{1\down}, \tbf{k}_{2\down})$
defined to be $(\tbf{k}, 2\tbf{K} -\tbf{k}, \tbf{k}, 2\tbf{K} - \tbf{k})$ and 
$(\tbf{k}, 2\tbf{K} - \tbf{k}, -2 \tbf{K} + \tbf{k}, -\tbf{k})$ for chiral charge and spin 
states, respectively. These $M_{\sigma \sigma'}$ matrices are 
\bea 
 &&  M_{\sigma \sigma} (\tbf{k}_{1\sigma}, \tbf{k}_{2 \sigma} ) \nn \\
&=&   \left[
		\begin{array}{cc} 
		\epsilon (\tbf{k}_{1\sigma} )  + \rho_\sigma U_{\sigma \sigma} (\tbf{K} - \tbf{k}) 
		       &	\rho_{\sigma} U_{\sigma \sigma} (\tbf{k} - \tbf{K})   \\
		\rho_\sigma U_{\sigma \sigma}^*  (\tbf{k} - \tbf{K}),	& \epsilon (\tbf{k}_{2\sigma})  
			   + \rho_\sigma U_{\sigma \sigma} (\tbf{K} - \tbf{k}) 
		\end{array}  
					\right]   \\ 
 &&  M_{\uparrow \downarrow} (\tbf{k}) \nn \\
& =& \left[ 
		\begin{array}{cc}
		U_{\up \down} (\tbf{k} - \tbf{K}) \sqrt{\rho_\uparrow \rho_\downarrow} & U_{\up \down} (\tbf{k} - \tbf{K}) \sqrt{\rho_\uparrow \rho_\downarrow} \\ 
		U_{\up \down} (\tbf{k} - \tbf{K}) \sqrt{\rho_\uparrow \rho_\downarrow} & U_{\up \down} (\tbf{k} - \tbf{K}) \sqrt{\rho_\uparrow \rho_\downarrow}   
		\end{array} 
	\right]   .
\eea 
From Bogoliubov Hamiltonian in Eq.~\eqref{eq:Hkmat}, we treat the spin-mixing part $M_{\uparrow \downarrow} $ 
as a perturbation which is well justified in the weakly interacting limit (see Fig.~\ref{fig:Esplit}a). 
We thus write $H_{\rm eff} = H_{\rm eff} ^{(0)} + H_{\rm eff} ^{(1)}$, with $H_{\rm eff} ^{(0)}$ block 
diagonal in the spin space. The leading part is readily diagonalized in terms of 
$\tilde{\Psi} (\tbf{k}) = 
[  \tilde{ \phi}_\uparrow (\tbf{k}_{+\uparrow} ), \tilde{\phi}_\uparrow ^\dag ( \tbf{k}_{-\up } ), 
\tilde{\phi}_\downarrow (\tbf{k}_{+\down}), \tilde{\phi}_\downarrow ^\dag  ( \tbf{k}_{-\down }) ]^T $, with 
\bea 
&& \tilde{\phi}_\sigma(\tbf{k}_{+,\sigma}) 
= u_\sigma (\tbf{k}_{1\sigma}, \tbf{k}_{2\sigma}) \phi_\sigma (\tbf{k}_{1\sigma} )
 + v_\sigma (\tbf{k}_{1\sigma}, \tbf{k}_{2\sigma}) \phi_\sigma ^\dag (\tbf{k}_{2\sigma} ), \nn \\ 
&& \tilde{\phi}_\sigma (\tbf{k}_{-,\sigma}) 
= v_\sigma  (\tbf{k}_{1\sigma}, \tbf{k}_{2\sigma})  \phi_\sigma (\tbf{k}_{1\sigma} ) 
+ u_\sigma (\tbf{k}_{1\sigma}, \tbf{k}_{2\sigma}) \phi_\sigma ^\dag (\tbf{k}_{2\sigma} ) .  
\eea 
The coefficients are determined to be 
\bea 
&& v_\sigma ^2  (\tbf{k}_{1\sigma}, \tbf{k}_{2\sigma})  = u_\sigma ^2  (\tbf{k}_{1\sigma}, \tbf{k}_{2\sigma})  -1  \nn \\
&&= \frac{1}{2} \left[ \frac{ \overline{\epsilon} (\tbf{k}_{1\sigma} , \tbf{k}_{2\sigma} ) + \rho_\sigma U_{\sigma \sigma} (\tbf{k} -\tbf{K})  }  
		      {\varepsilon_\sigma (\tbf{k}_{1\sigma}, \tbf{k}_{2\sigma} )}
		    -1 \right], \nn 
\eea 
with 
\bea 
\varepsilon_\sigma ^2  (\tbf{k}_{1\sigma}, \tbf{k}_{2\sigma})  
	&=&  \overline{\epsilon} (\tbf{k}_{1\sigma}, \tbf{k}_{2\sigma} ) 
	    \left[ \overline{\epsilon} (\tbf{k}_{1\sigma}, \tbf{k}_{2\sigma}) + 2\rho_\sigma U_{\sigma \sigma} (\tbf{K} -\tbf{k})  \right]\nn \\ 
\overline{\epsilon} ( \tbf{k}_{1\sigma} , \tbf{k}_{2\sigma} ) &=& 
\left( \epsilon(\tbf{k}_{1\sigma}) + \epsilon (\tbf{k}_{2\sigma} ) \right) /{2}.\nn 
\eea 
The Bogoliubov spectra are 
\bea
\xi_{\pm , \sigma} 
	=  \varepsilon_\sigma (\tbf{k}_{1\sigma}, \tbf{k}_{2\sigma}) \pm \Delta \epsilon (\tbf{k}_{1\sigma} , \tbf{k}_{2\sigma}) 
\eea 
with 
$$
\Delta \epsilon (\tbf{k}_{1\sigma} , \tbf{k}_{2\sigma}) =(\epsilon(\tbf{k}_{1\sigma}) - \epsilon(\tbf{k}_{2\sigma} ))/{2}.
$$
Under the condition $U_{\sigma \sigma} (\tbf{k}) >0$ already assumed , we have $\xi_{\pm, \sigma} >0$, 
which means the system is stable~\cite{2013_Stamper_Kurn_RMP,2008_Pethick_Smith}. 
Then $H_{\rm eff}^{(0)}$ takes a diagonal form  
$
H_{\rm eff} ^{(0)} = \frac{1}{2}\int_\tbf{k} 
	\sum_p \xi_{p ,\sigma}  \tilde{\phi}_{\sigma} ^\dag (\tbf{k}_{p, \sigma})   \tilde{\phi}_{\sigma} (\tbf{k}_{p, \sigma}) 
+ { E} ^{(0)}, 
$ 
with ${ E} ^{(0)}$, 
\bea 
&& \textstyle  { E} ^{(0)}/N_s \nn \\
&=&\textstyle \frac{1}{2} \int _\tbf{k} 
\left\{  -2 \bar{\epsilon} (\tbf{k}, 2\tbf{K} - \tbf{k}) +  
     \sum_\sigma 
	 \varepsilon_\sigma (\tbf{k}, 2\tbf{K} - \tbf{k}) 
	- \rho_\sigma   U_{\sigma \sigma} (\tbf{K} - \tbf{k})  
\right\}  
\eea 
the same for chiral charge and spin states.

Treating the spin mixing terms perturbatively, (see Sec.~\ref{sec:perttheory}) the ground state receives an energy correction 
\bea 
 \textstyle  {E}^{(2)}/N_s 
\textstyle & =&\textstyle -\frac{1}{2} {\rho_\up \rho_\down} \int  _\tbf{k}      
     \Gamma^2 (\tbf{k}_{1\up}, \tbf{k}_{2\up}, \tbf{k}_{1\down},\tbf{k}_{2\down} ) \nn \\
&& \times  \textstyle    \left\{\frac{1 }{\xi_{+\up} + \xi_{- \down} } 
      + \frac{1 }{\xi_{+ \down} + \xi_{-\up}}   \right\}, 
\eea 
with 
\bea  
&& \Gamma (\tbf{k}_{1} , \tbf{k}_{2}, \tbf{k}_{3},\tbf{k}_{4} ) \nn \\
&=&  \textstyle | U_{\up \down} ( \tbf{k} - \tbf{K}) |^2
 \left(u_\up (\tbf{k}_{1}, \tbf{k}_{2} ) - v_\up (\tbf{k}_{1}, \tbf{k}_{2} ) \right)
      \left( u_\down (\tbf{k}_{3}, \tbf{k}_{4} ) - v_\down (\tbf{k}_{3}, \tbf{k}_{4}) \right) . 
\eea  
Introducing 
\be 
g(\tbf{k}) = \Gamma (\tbf{k}, 2\tbf{K} -\tbf{k}, \tbf{k}, 2\tbf{K} -\tbf{k} ),
\ee 
for the chiral charge case we have 
\bea 
\textstyle { E}_{\chi_c} ^{(2)} /N_s = 
\textstyle -\frac{1}{2} {\rho_\up \rho_\down}
\int _\tbf{k}   g^2(\tbf{k} )  
\times  \textstyle 
    \left\{  \frac{2 } {\varepsilon_\up (\tbf{k}, \tbf{Q} -\tbf{k}) 
			  + \varepsilon_\down(\tbf{k}, \tbf{Q} - \tbf{k}) } 
		 \right\}, 
\label{eq:Echic} 
\eea 
with $\tbf{Q} = 2\tbf{K}$, 
while for the chiral spin case we have 
\bea 
&& \textstyle {E}_{\chi_s} ^{(2)} /N_s = -\frac{1}{2} {\rho_\up \rho_\down} 
  \int  _\tbf{k}  g^2 (\tbf{k}) \nn \\
&  \textstyle  \times & \textstyle 
  \left\{  \frac{1}{\varepsilon_\up (\tbf{k}, \tbf{Q} -\tbf{k}) 
			  + \varepsilon_\down(-\tbf{Q} + \tbf{k}, -\tbf{k} )
			    + \Delta \epsilon  (\tbf{k},  \tbf{Q} - \tbf{k}) 
			    - \Delta \epsilon (-\tbf{Q} + \tbf{k}, -\tbf{k})  } \right. \nn \\
&& \left. \textstyle	
  + \frac{1}{\varepsilon_\down (-\tbf{Q} +\tbf{k}, -\tbf{k}) 
		    + \varepsilon_\up (\tbf{k}, \tbf{Q} -\tbf{k}) 
		    + \Delta \epsilon (-\tbf{Q} + \tbf{k}, -\tbf{k}) 
		    - \Delta \epsilon (\tbf{k},  \tbf{Q} - \tbf{k} ) } \right\}. 
\label{eq:Echis} 
\eea 
 
We also calculate the energy correction by numerically diagonalizing the full Bogoliubov Hamiltonian (Eq.~\eqref{eq:Hkmat}), finding 
excellent agreement with our analytic results when the inter-species interactions are weak (see Fig.~\ref{fig:Esplit}a). 

\begin{figure}[htp]
\includegraphics[angle=0,width=0.5\linewidth]{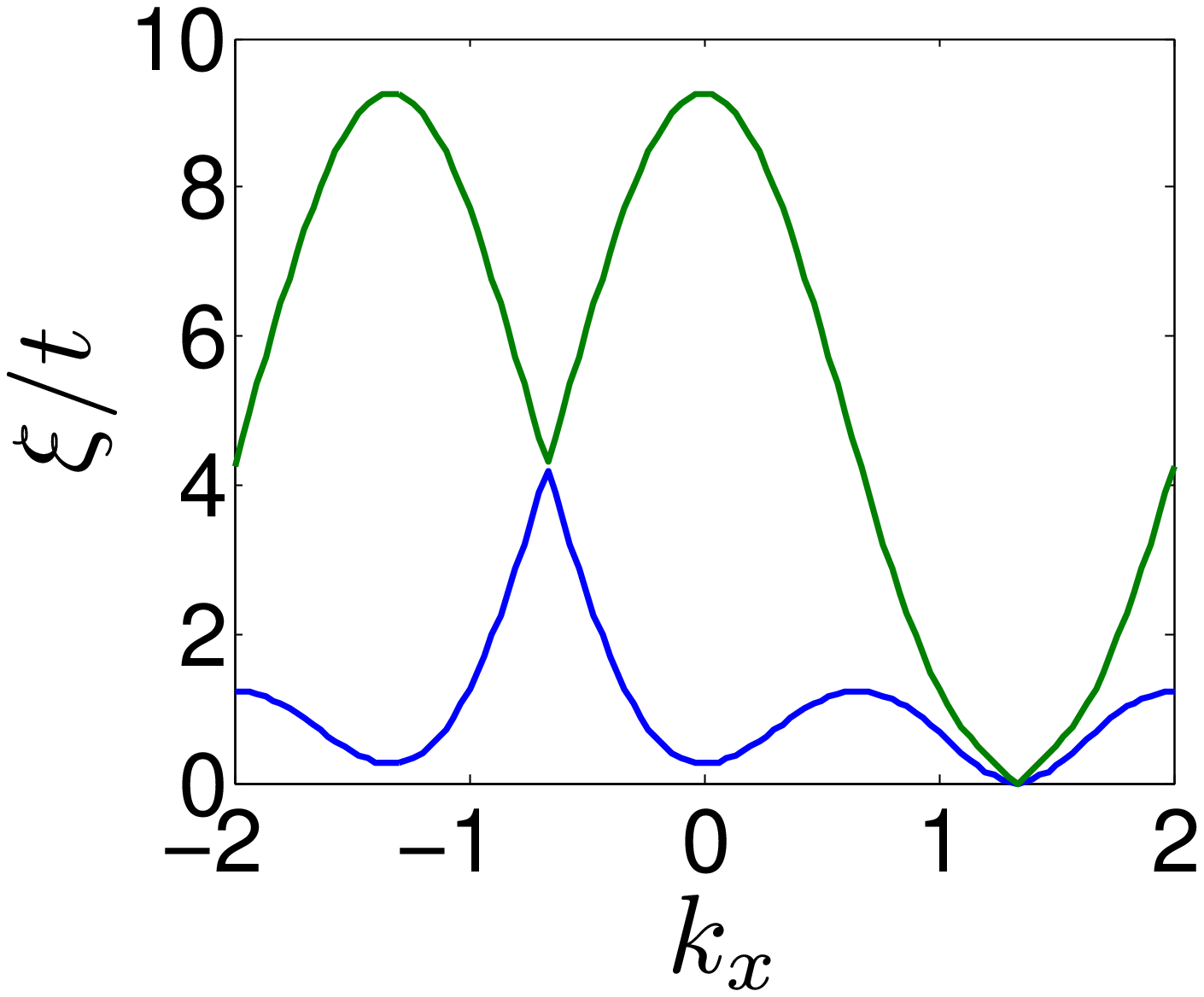}
\caption{Bogoliubov spectra  along the $k_x$ axis in the chiral spin superfluid state. In this plot we use $U/t = 1$ and $V/t = 0.3$. } 
\label{fig:bogspectra} 
\end{figure}

\section{Perturbation theory for Bogoliubov ground states}
\label{sec:perttheory} 
In this section, we discuss the perturbative method to calculate the ground state energy of a  Bogoliubov problem 
$$
H_{\rm Bog}  = \Psi^\dag {\cal H}_{\rm Bog}  \Psi - {\cal H} _{\rm Bog} (2,2) - {\cal H}_{\rm Bog} (4,4), 
$$ 
with $\Psi$ a column vector of bosonic operators 
$[\phi_{\up 1}, \phi_{\up 2} ^\dag, \phi_{\down 1}, \phi_{\down 2} ^\dag] ^T$. 
This Bogoliubov Hamiltonian is one momentum slice of Eq.~\eqref{eq:SHeff} 
and the momentum ${\bf k}$ index is suppressed for brevity. 
The $4\times4$ matrix ${\cal H}_{\rm Bog}$ can be rewritten as 
\bea 
{\cal H}_{\rm Bog}  = \left[ 
	    \begin{array}{cc} 
	      M_\up 	&G  \\ 
	      G^\dag 	&M_\down 
	       \end{array}
	      \right], 
\eea 
where the $2\times 2$ matrices can be expanded in terms of Pauli matrices, 
$M_\sigma = c_{0\sigma}  \mathbb{1} + c_{x\sigma} \sigma_x + c_{z \sigma} \sigma_z$, and 
$G$ takes a special form  $g (\mathbb{1} + \sigma_x)$. The terms $c_{0\sigma}$, $c_{x\sigma}$ and $c_{z\sigma}$ can be read off from Eq.~\eqref{eq:Hkmat}.
Here we will treat the off-diagonal part $G$ perturbatively. 
The leading part is readily diagonalized as 
\bea 
\textstyle H^{(0)}  =\sum_ \sigma \textstyle \left[ 
       \tilde{\phi}_{\sigma + }^\dag , \tilde{\phi} _{\sigma -} 
      \right] 
      D_\sigma   
      \left[ 
	\begin{array}{c} 
	 \tilde{\phi}_{\sigma + } \\ 
	  \tilde{\phi} _{\sigma -}^\dag  
	\end{array} 
      \right] 
+ D_\sigma (2,2) - M_\sigma (2,2) \nn 
\eea 
with 
\bea 
D_\sigma &=& \epsilon_\sigma + c_{z\sigma} \sigma_z \\ 
\epsilon_\sigma &=& \sqrt{c_{0\sigma}^2 - c_{x \sigma}^2} \nn 
\eea 
and 
\bea 
\left[ 
  \begin{array}{c} 
   \tilde{ \phi}_{\sigma +} \\ 
   \tilde{\phi} _{\sigma -} ^\dag 
  \end{array}
  \right] = T_\sigma 
\left[ 
  \begin{array}{c} 
   { \phi}_{\sigma 1} \\ 
   {\phi} _{\sigma 2} ^\dag 
  \end{array}
  \right], \nn  
\eea 
\bea 
T_\sigma = 
\left[ 
\begin{array}{cc} 
u_\sigma	&v_\sigma \\ 
v_\sigma	&u_\sigma 
\end{array} 
\right] \nn 
\eea 
and 
\be 
u_\sigma^2 = v_\sigma ^2 + 1 = \frac{1}{2}\left[ \frac{c_{0\sigma}}{\epsilon_\sigma} +1\right]. 
\ee 
The Bogoliubov spectra are 
\bea 
\xi_{\sigma, \pm} = \epsilon_\sigma \pm c_{z \sigma} 
\eea 
In terms of $\tilde{\phi}$, 
the perturbative part reads 
\bea 
&& H^{(1)} = 
g(u_\up - v_\up ) (u_\down - v_\down) \nn \\
&\times& \textstyle 
\left\{ 
  \tilde{ \phi}_{\up + } ^\dag \tilde{\phi}_{\down +}  + \tilde{\phi}_{\up -} \tilde{\phi}_{\down -} ^\dag 
  + \tilde{\phi} _{\up - } \tilde{\phi}_{\down +} + \tilde{\phi}_{\up +} ^\dag \tilde{\phi}_{\down -} ^\dag 
\right\} + h.c. 
\label{eq:SH1}
\eea 
Then  standard perturbation theory applies, and only the third and fourth terms in Eq.~\eqref{eq:SH1} contribute at second order.  
The ground state energy is thus obtained to be 
\bea 
 && { E}  
=\sum_\sigma (\epsilon_\sigma -  c_{0, \sigma} )   \\
&-&  \left|g(u_\up - v_\up) (u_\down -v_\down) \right|^2 
 \times  \left\{
  \frac{1}{\xi_{\up +} + \xi_{\down -} } 
  + \frac{1}{\xi_{\up +} + \xi_{\down -} } \right\}.  \nn  
\eea

\section{Large $U$ limit of the hexagonal lattice model}
\label{sec:largeU} 
In the large $U$ limit, we can project out double occupancy, and the Gutzwiller state is
\bea 
|G \rangle  
= \prod_\tbf{r} \left( f_{\up, \tbf{r},0} + f_{\up, \tbf{r}, 1} \phi_{\up, \tbf{r}} ^\dag  \right) 
	      \left( f_{\down, \tbf{r}, 0} + f_{\down, \tbf{r}+\hat{e}_1, 1} \phi_{\down, \tbf{r}+\hat{e}_1} ^\dag \right) 
	      |{\rm vac} \rangle, \nn 
\eea 
with a normalization condition $|f_{\sigma, \tbf{r}, 0}|^2 + |f_{\sigma, \tbf{r}, 1}|^2  =1$. 
To minimize kinetic energy we take $f_{\up/\down, \tbf{r}, 1} = f_{\up/\down,1} e^{\pm i\tbf{K} \cdot \tbf{r}}$ 
and $f_{\sigma, \tbf{r}, 0} = \sqrt{1-f_{\sigma, 1}^2}$, where $f_{\sigma,1}$ is a  real number. Then the energy 
cost of the Gutzwiller state is 
\bea 
&& E /N_s  =  \left[-\mu -3t_{\rm eff} \right] \left( f_{1\up} ^2 + f_{1\down}^2  \right)  \\ 
	&+& \frac{3t_{\rm eff} + 3V/2}{2}  \left( f_{1\up} ^2 + f_{1\down}^2  \right) ^2 
	+ \frac{3t_{\rm eff}  -3V/2} {2} \left( f_{1\up} ^2 - f_{1\down}^2  \right) ^2 ,  \nn 
\eea 
which after minimization leads to 
\bea
\left\{ 
\begin{array}{cccc} 
| f_{1\up} ^2 - f_{1\down} ^2|&=&0 , &\text{if $t_{\rm eff} >V/2$} \\ 
| f_{1\up}^2  - f_{1\down} ^2|& =& \frac{2\mu + 6t_{\rm eff} }{6 t_{\rm eff} + 3V}, &\text{otherwise} 
\end{array}
\right. .
\eea 
The transition from the  unpolarized superfluid to the fully polarized state is at $V_c = 2t_{\rm eff}$ in this large $U$ limit, 
where the transition is first order.

\end{document}